# Artificial double-helix for geometrical control of magnetic chirality

Dédalo Sanz-Hernández[1,#,*], Aurelio Hierro-Rodriguez[2], Claire Donnelly[1], Javier Pablo-Navarro[3], Andrea Sorrentino[4], Eva Pereiro[4], César Magén[3,5], Stephen McVitie[2], José María de Teresa[3,5], Salvador Ferrer[4], Peter Fischer[6,7], Amalio Fernández-Pacheco[1,2,*].

[1]Cavendish Laboratory, University of Cambridge, JJ Thomson Cambridge, CB3 0HE, United Kingdom
[2]SUPA, School of Physics and Astronomy, University of Glasgow G12 8QQ, UK.
[3]LMA, Instituto de Nanociencia de Aragón, Universidad de Zaragoza, 50018 Zaragoza, Spain
[4]ALBA Synchrotron, 08290 Cerdanyola del Vallès, Spain.
[5]Instituto de Ciencia de Materiales de Aragón (ICMA), Universidad de Zaragoza-CSIC, 50009 Zaragoza, Spain
[6]Materials Sciences Division, Lawrence Berkeley National Laboratory, Berkeley, CA 94720, USA
[7]Physics Department, University of California Santa Cruz, Santa Cruz, CA 95064, USA
[#]Now at Unité Mixte de Physique, CNRS, Thales, Université Paris-Saclay, 91767, Palaiseau, France
*Corresponding authors: dedalo.sanz@cnrs-thales.fr, amalio.fernandez-pacheco@glasgow.ac.uk

**Chirality plays a major role in nature, from particle physics[1] to DNA[2], and its control is much sought-after due to the scientific and technological opportunities it unlocks[3–5]. For magnetic materials, chiral interactions between spins promote the formation of sophisticated swirling magnetic states such as skyrmions[6,7], with rich topological properties and great potential for future technologies[8]. Currently, chiral magnetism requires either a restricted group of natural materials[6,7] or synthetic thin-film systems that exploit interfacial effects[9–11]. Here, using state-of-the-art nanofabrication[12,13] and magnetic X-ray microscopy[14–17], we demonstrate the imprinting of complex chiral spin states via three-dimensional geometric effects at the nanoscale. By balancing dipolar and exchange interactions in an artificial ferromagnetic double-helix nanostructure, we create magnetic domains and domain walls with a well-defined spin chirality, determined solely by the chiral geometry. We further demonstrate the ability to create confined 3D spin textures and topological defects by locally interfacing geometries of opposite chirality. The ability to create chiral spin textures via 3D nano-patterning alone enables exquisite control over the properties and location of complex topological magnetic states, of great importance for the development of future metamaterials and devices in which chirality provides enhanced functionality.**

Chirality, or "handedness", refers to the property characterising three-dimensional structures that cannot be superimposed on their mirror image, as defined by Kelvin[18], Larmor[19] and Eddington[20]. In nature, chirality is critical in determining a material's functionality, from the chirality of molecules[21] dictating key differences in properties such as flavour, toxicity, or drug effectiveness, to the catalysis of important biological reactions that played a role in the origin of life[22]. Chirality is a fundamentally three-dimensional property, that until recently was observed mostly in naturally occurring materials. With significant developments in nanofabrication, it has recently become possible to *design chirality* in metamaterials and devices. In particular, by patterning materials at the micro- and nano-scales in chiral geometries, broadband optical polarisers[23], micro-robots[24] and sensors[25] have been demonstrated.

In magnetism, chirality plays a critical role in the stabilisation of chiral domain walls[26] and spin spirals[27], as well as in the creation of topological spin textures such as skyrmions[6,7], antiskyrmions[28] and bobbers[29], which are very attractive for future non-volatile computing technologies[8]. In recent years, the antisymmetric exchange Dzyaloshinskii–Moriya interaction (DMI)[30,31] describing chiral spin coupling, has been achieved either in certain non-centrosymmetric bulk crystals[6,7], or through the careful design of high quality layered thin films and interfaces[9–11], leading to a fascinating array of magnetic effects[32–35].

An alternative route to achieve magnetochirality is through 3D patterning. Relying solely on shape, geometric magnetochirality eliminates the stringent material requirements associated with bulk and interfacial DMI. In this realm, single-domain chiral magnetic metamaterials have been realised[36], and a range of magnetochiral effects have been predicted for inversion symmetry-breaking curved nanostructures[37,38], with recent experiments revealing geometric DMI to be of comparable magnitude to interfacial DMI[39].

Here, via the combination of advanced 3D nanofabrication and magnetic microscopy, we reveal a rich and sophisticated collection of magnetic states reachable by purely 3D geometric means, ranging from helical domains to chiral spin textures and localised topological defects. As a prototypical system, we have designed a magnetic double-helix formed of two twisted and overlapping cylindrical nanowires (**Fig. 1**). The double helix combines dipolar (**Fig. 1a**) and exchange (**Fig. 1b**) coupling with geometrically-induced chirality (**Fig. 1c**), giving rise to a rich magnetic energy landscape.

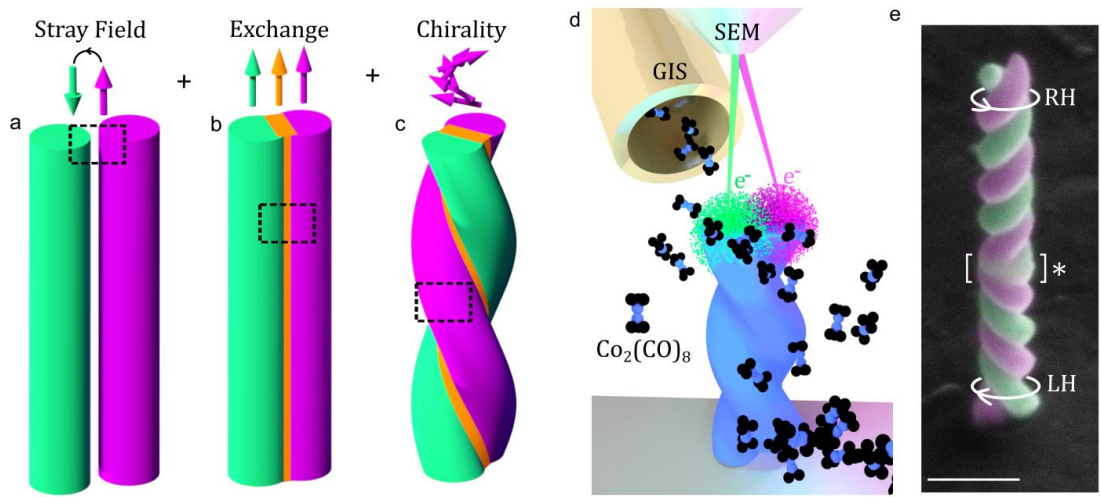

*Figure 1 - Artificial double-helix nanomagnets. (a-c) Competing magnetic interactions in an artificial double-helix. Colours separate different regions of the system (two strands in green and magenta, and core in yellow), with no difference in material composition. Arrow insets indicate spin alignment inside the dashed boxes. (a) Dipolar stray field minimisation promotes an antiparallel state between strands. (b) Exchange coupling in the overlap region (yellow) favours the parallel alignment of spins. (c) Geometric chirality favours a unique sense of spin rotation in a strand via shape anisotropy. (d) 3D-printing of a Cobalt nano-helix by FEBID. After injection of $Co_2(CO)_8$ into the chamber of a scanning electron microscope (SEM) using a gas injection system (GIS), the focused electron beam (in green and magenta) alternatively exposes the two helix strands. (e) Coloured SEM image of the nanostructure under investigation, consisting of two double-helices of opposite chirality joined at the tendril perversion marked ∗. Scale bar 250nm, image tilt 45°.*

To experimentally realise this system, we have 3D-printed cobalt double-helix nanomagnets using focused electron beam induced deposition (FEBID), as illustrated in **Fig. 1d**. The unique performance of FEBID for the direct-writing of ferromagnetic materials[13,40] is exploited here to not only create double-helices with a single chirality, but also to combine two of them with opposite chirality (**Fig. 1e**): left-handed (LH) and right-handed (RH) in the lower and upper parts, respectively. These two regions of opposite geometric chirality are interfaced via a structural defect, a nanometric tendril perversion[41] (see ∗ in **Fig. 1e**). Each of the two nanowires forming the double-helix strands has a total length of 880 nm and a diameter of 85 nm, favouring an axial magnetic domain configuration due to shape anisotropy[42]. Transmission electron microscopy (TEM) measurements reveal a geometrical overlap of 18 nm between the two strands, resulting in strong inter-strand ferromagnetic exchange coupling. Further details are available in Methods.

In order to identify the magnetic structure emerging in these systems, a high spatial resolution magnetic imaging technique is required[17,36,43–45]. Here, we use full-field magnetic soft X-ray transmission microscopy[14–16] (**Fig. 2a**), exploiting X-ray magnetic circular dichroism (XMCD) at the Cobalt $L_3$ absorption edge. Using this technique, it is possible to infer the projection of magnetic moments parallel to the X-ray beam (x-direction in **Fig. 2a**), by comparing two calibrated images taken with different X-ray polarisation. More details are available in Methods.

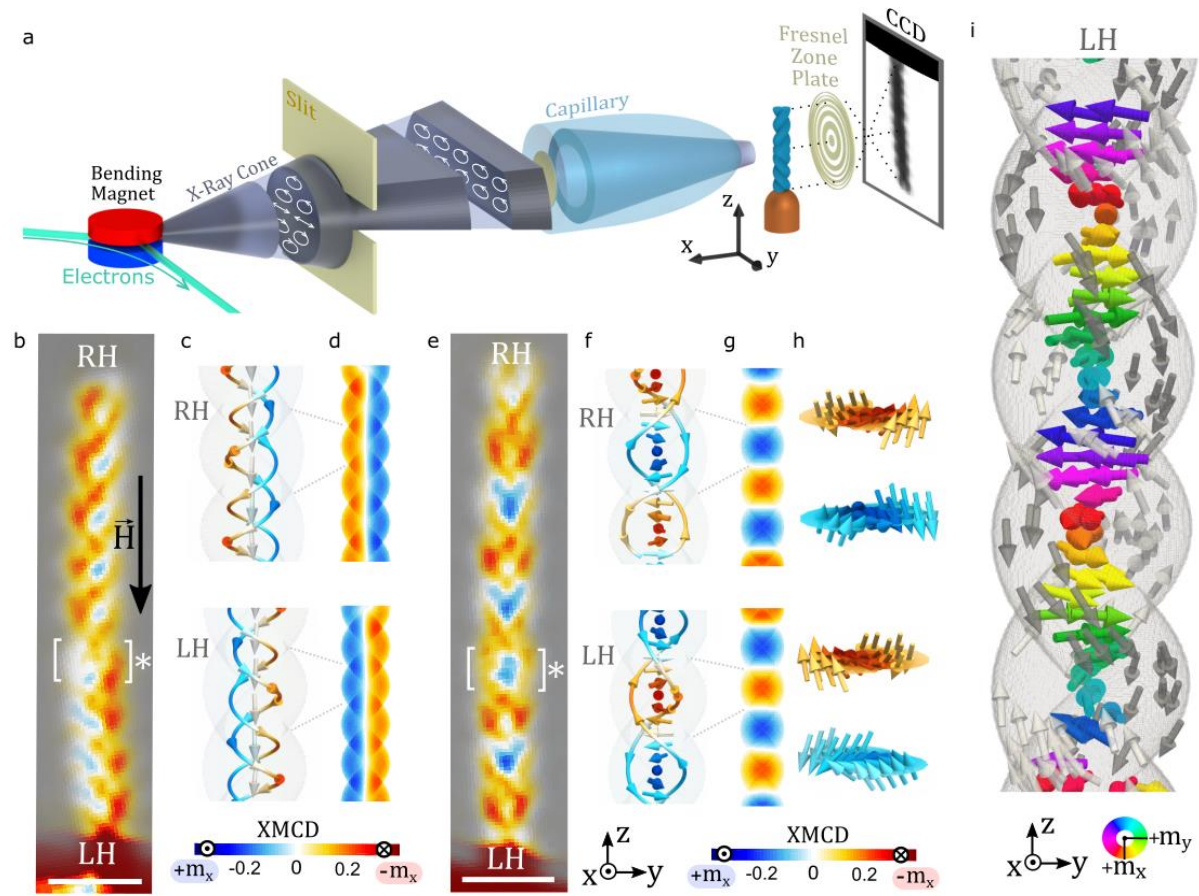

*Figure 2 – Geometric imprinting of magnetic chirality. (a) Schematic setup for X-ray magnetic microscopy. X-rays are focused by a capillary condenser onto the sample. A magnified image is formed on a CCD by a Fresnel Zone plate. Two exposures with different polarisation (selected by a slit) are taken to extract X-ray Magnetic Circular Dichroism (XMCD) contrast, which is proportional to the x component of the nanostructure's magnetisation. (b) XMCD image of the double-helix studied, which changes geometric chirality at ∗. Image at zero field, after application of a saturating field along -z. (c-d) Micromagnetic simulations of a RH (top) and LH (bottom) double-helix, after application of a saturating external field along -z. (c) Magnetic state for strands and core. (d) Computed XMCD signal from simulations in (c). (e) XMCD image of the double-helix under study in the as-grown state. (f-h) Micromagnetic simulations for a RH (top) and LH (bottom) double helix with antiparallel magnetic alignment of the strands. (f) Internal spin structure for strands and core (g) Corresponding calculated XMCD contrast. (h) Selected xy cross-sections at different heights, revealing a chiral inter-strand Bloch domain wall. (i) Volumetric representation of the 3D spin configuration of a single LH double helix with antiparallel magnetic alignment of the strands (grey arrows) and a helical Bloch domain wall in the core (colour arrows). Scale bars (b, e): 200nm.*

We image the magnetic state of the sample in two different remanent states: after applying an axial (> 0.5 T) magnetic field (**Figs. 2b-d**), and the as-grown magnetic configuration (**Figs. 2e-i**). After the application of the axial magnetic field (**Fig. 2b**), a mirrored XMCD contrast is observed for the regions with opposite geometric chirality of the double helix: In the LH bottom part of the structure, the left-hand side has a positive (blue) XMCD signal, contrary to the right-hand side, where it is negative (red). This indicates that the projection of the magnetisation along the *x* direction reverses when moving from the left to the right side of the structure. The opposite XMCD contrast is observed for the RH upper section of the helix. Such a pattern is consistent with the two intertwined helices being longitudinally magnetised in the direction of the applied field (parallel strand configuration). Due to the overlap between strands, the resulting magnetic configuration and XMCD contrast resembles a vortex state, as *e.g.* in cylindrical nanowires of large diameter[46]. Importantly, the magnetic circulation is opposite for LH and RH regions, *i.e.* it is dependent on the geometric chirality.

To elucidate details of this magnetic configuration, we have performed micromagnetic simulations[47] at remanence, after a saturating field along the helix axis (**Fig. 2c**). The state is characterised by a shell of helical spins following the strand axes, due to dominant magnetic shape anisotropy. This results in the imprinting of magnetic LH (RH) helical domains at the bottom (top) parts of the structure, with a magnetic core parallel to the applied magnetic field (-z direction). The corresponding XMCD image calculated from the micromagnetic state is shown in **Fig. 2d**, where the key feature observed in experiments, *i.e.* reversal of blue-red contrast for a change in geometric chirality, is reproduced. This result demonstrates the ability to directly imprint magnetic chiral states by exploiting the shape of an artificial helical nanomagnet.

We now explore the *as grown* magnetic state of the double-helix, revealing a rich magnetic configuration. The experimental XMCD image (**Fig. 2e**) consists of a longitudinally alternating contrast, that appears independent of the geometric handedness (either LH or RH) of the double-helix. This alternating magnetic contrast is consistent with an antiparallel magnetic alignment of the strands, as the micromagnetic simulations (**Figs. 2f**) and computed XMCD contrast (**Fig. 2g**) elucidate. Small distortions in the experimental projection with respect to the simulated signal are well reproduced by considering the effect of small changes in illumination angles between both polarisations (see Methods). Simulations also indicate that the antiparallel strand alignment experimentally observed forms the ground state of the system. This state is expected to be formed during growth due to the high dipolar coupling between strands (**Fig. 1a**), which favour the opposite $m_z$ magnetic configuration of the two strands (see Methods).

An intriguing magnetic structure becomes apparent when we consider the internal spin arrangement of this antiparallel magnetic state, as a closer look at the micromagnetic configuration reveals. Due to the existence of antiparallel magnetic domains for the two strands, a domain-wall at the core of the double-helix must form (see core spins in **Fig. 2f** and in **Fig. 2i** for greater detail). Considering the spin evolution of the wall from bottom to top, a helical pattern is formed, with an opposite helicity for the LH (bottom) and RH (top) regions (**Fig. 2f**). The chirality of this 3D spin wall is fully determined by the geometric chirality of each part of the structure. Moreover, if we examine the sense of rotation of the magnetisation during the transition between strands (see cross sections along the xy plane in **Fig. 2h**), this takes place via a Bloch domain wall with a well-defined chirality. This Bloch wall chirality is fundamentally determined by the non-zero angle formed between the strands, being thus opposite for top and bottom parts of the structure. This result constitutes a striking manifestation of the imprinting of geometric chirality into the magnetic texture of the double-helix.

We have demonstrated how the sophisticated patterning of a symmetry-breaking 3D nano-geometry translates into a powerful control of magnetic chirality. In the following, we study in more detail the type of 3D spin state that emerges when two regions with opposite magnetic chirality are forced to interact at the inter-chiral interface (∗). Specifically, we focus on the antiparallel strand configuration, for which two helical Bloch walls of opposite chirality meet at (∗). To determine the 3D magnetic texture present at this chirality interface, the double-helix is rotated around its axis, as shown schematically in **Fig. 3a**, and XMCD projections are measured at different rotation angles. An electronic contrast X-ray transmission image is shown in **Fig. 3b**, where each half pitch of the double-helix is identified I to IX, with the perversion at turn IV∗. **Fig. 3c** and **Fig. 3d** correspond to the simulated and experimental XMCD images at different angles, respectively. We first note that the XMCD contrast pattern shifts upwards (downwards) for the RH (LH) helix by half a period upon rotation by 90°, similarly to a barber pole. This confirms the antiparallel magnetic configuration. A significant change in the magnetic contrast is additionally observed at the chirality interface (∗) during rotation, changing smoothly from a blue dot at 0° to a horizontally split red-blue contrast for the highest angles probed. This magnetic contrast evolution is reminiscent of the one corresponding to a magnetic vortex: for a vortex, in the perpendicular geometry the core is probed, which is seen as a bright or dark point; and as the sample is rotated into the in-plane geometry, the planar magnetic domains are probed, which exhibit opposite XMCD contrast at either side of the core. Micromagnetic simulations of the perversion (**Fig. 3e**) reveal that indeed, a 3D vortex-like spin structure is stable at the perversion, and results in XMCD projections (**Fig. 3d**) that match well the experimentally-observed contrast (**Fig. 3c**).

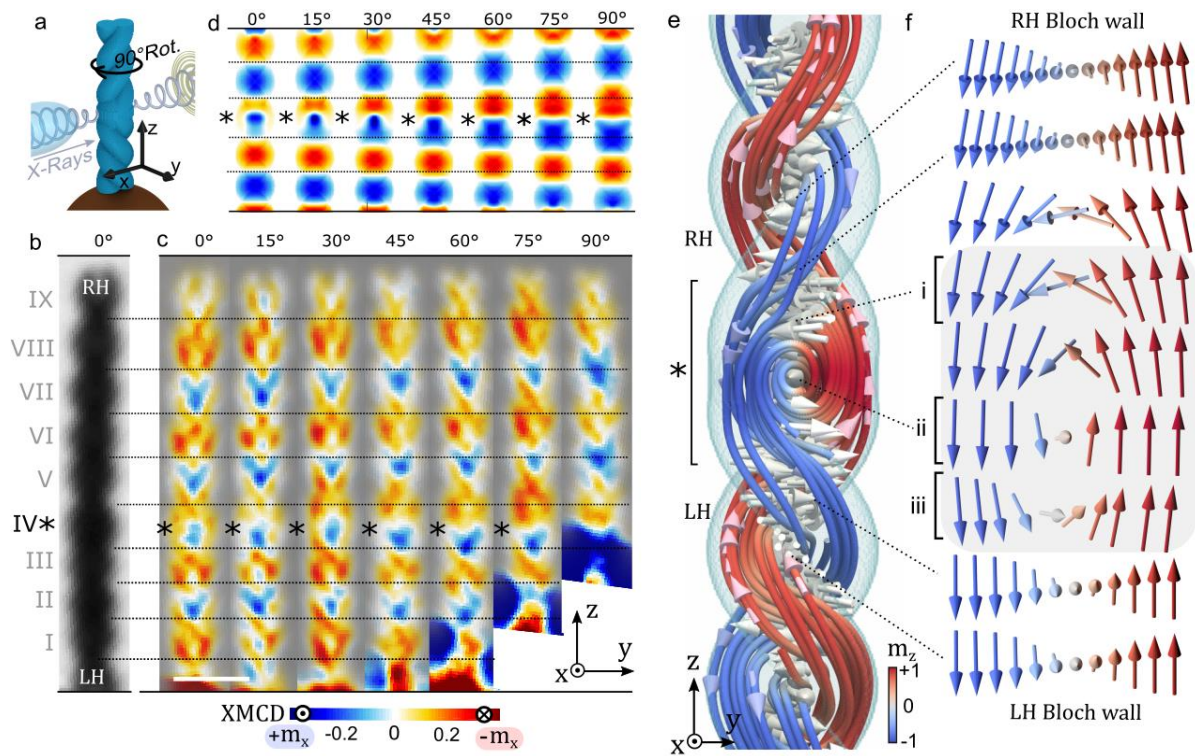

*Figure 3 – Localised 3D spin texture and topological defect in a chirality interface. (a) Schematics of the experimental procedure, by which the double-helix is rotated to measure X-ray magnetic images at different angles. (b) Reference electronic-contrast X-ray transmission image. Each double-helix half-pitch is marked from I to IX (tendril perversion located at IV∗). (c) XMCD measurements taken at several rotation angles. A vortex-like spin state is identified from the angular evolution of the magnetic contrast in region ∗. (d) Simulated XMCD contrast for the spin structure in (e), reproducing the magnetic pattern experimentally observed upon rotation. (e) Micromagnetic simulations of the linkage between two double-helices of opposite chirality with antiparallel magnetic alignment of their strands (blue and red arrows). LH (bottom) and RH (top) Bloch domain walls are present in the core (white arrows), and an asymmetric 3D vortex state emerges at the tendril perversion ∗. Represented viewpoint corresponds to 0° measurements. (f) Subset of spins in (e) after untwisting the magnetisation state. Line cross-sections are taken in-between the centres of the two helix strands at different heights and rotated into the xy plane revealing that the change in magnetic chirality (transition from a RH to a LH Bloch wall) takes place via the formation of an asymmetric vortex (grey area). The Néel-type defect mediating opposite chiralities (i), the vortex core (ii) and the deformed Bloch-wall (iii) discussed in the main text are indicated. Scale bar in (c): 200 nm.*

The micromagnetic configuration of the 3D vortex-like spin state can be further understood by untwisting the helical Bloch domain wall into a plane, as shown in **Fig. 3f**. This procedure reveals that when two Bloch domain-walls of opposite chirality meet at the perversion, an inversion of magnetic chirality occurs via the transformation of the Bloch walls to an intermediate Néel wall, indicated by (i) in **Fig. 3f**. The emergence of this topological defect is analogous to the formation of Bloch lines in materials hosting magnetic bubbles, where a line with Néel character arises to mediate a chirality transition between opposite-chirality Bloch domain walls[15]. Here, the presence of this Néel defect generates a circulation in the magnetisation around a central core (ii), as detected experimentally. The resulting spin configuration, however, is not equivalent to a traditional vortex found in planar systems, where a symmetric circulation with respect to the vortex core would take place. An asymmetric 3D vortex state arises instead, characterised by the deformation of the LH Bloch domain wall below the core (iii), an effect promoted by the geometry and dimensions of the perversion, in combination with the twisting of the interconnected strands. As well as the asymmetric vortex with RH chirality observed experimentally at the perversion, micromagnetic simulations reveal that other states such as a Bloch-point wall may also be stable (Methods). These results establish the inter-chirality boundary as a powerful platform for the formation of complex 3D magnetic states.

This work demonstrates an unprecedented control over spin chiral properties via purely three-dimensional geometric nanoscale effects. By a careful design of the geometry of a three-dimensional chiral nanomagnet, patterned via advanced direct-write lithography, we are able to imprint sophisticated chiral spin states. We control the helicity and location of magnetic domains and domain walls, and the internal chirality of domain walls interfacing helical domains. We also show how via the introduction of structural defects mediating opposite magnetochiral regions, localised 3D spin textures and topological defects can be realised. The platform shown here represents an exciting route to study three-dimensional chiral spin systems, discover novel spin textures[48,49] and develop new magnetic nano-devices[50].

# Acknowledgements

This work was funded by EPSRC Early Career Fellowship EP/M008517/1, the Winton Program for the Physics of Sustainability and the EU CELINA COST action. DSH acknowledges a Girton College Pfeiffer scholarship and support from the EPSRC CDT in Nanoscience and Nanotechnology. AHR and SMV acknowledge funding from the EU Horizon 2020 program through Marie Skłodowska-Curie Action H2020-MSCA-IF-2016-74695. CD acknowledges funding from Leverhulme Trust (ECF-2018-016), Isaac Newton Trust (18-08) and a L'Oréal-UNESCO UK and Ireland Fellowship for Women in Science 2019. Funding by the Spanish Ministry of Science is acknowledged, grants MAT2017-82970-C2-1-R, MAT2017-82970-C2-2-R and MAT2018-102627-T, and by Aragon Government (Construyendo Europa desde Aragón), grant E13_17R including European Social Fund. JPN acknowledges MINECO funding BES-2015-072950. SMV thanks support from EPSRC EP/M024423/1. PF was supported by the U.S. Department of Energy, Office of Science, Office of Basic Energy Sciences, Materials Sciences and Engineering Division, Contract No. DE-AC02-05-CH11231 (NEMM program MSMAG). We would like to thank Robert Streubel, Mi-Young Im and Máximo Sanz-Hernández for stimulating discussions. We acknowledge William Smith, Colin How and Samuel McFadzean for electron microscopy technical support and Ricardo Valcarcel for X-ray microscopy technical support.

# Author contributions
DSH and AFP designed the experiments. DSH and AHR fabricated the samples. DSH performed micromagnetic simulations. DSH, AHR, CD, AS, JPN, SF, PF and AFP performed X-ray Microscopy experiments. DSH developed X-ray measurement protocols with the help of CD, AHR and AFP. DSH, CD and AHR analysed the X-ray data. E.P. built, maintained, and supervised the X-ray experiment. AHR performed TEM measurements. DSH, AHR, CD and AFP wrote the paper. DHS created the figures. All authors contributed to the interpretation and discussion of results.

# Data availability
Raw data, Mumax3 simulation scripts and simulation results are available publicly at Enlighten, the data repository of the University of Glasgow. The code employed to calculate XMCD images from micromagnetic data is available from the corresponding authors upon reasonable request.

# Competing interests
The authors declare no competing interests.